\documentclass[prd,aps,twocolumn,nofootinbib,preprintnumbers,superscriptaddress,preprintnumbers,balancelastpage,longbibliography]{revtex4-1}
\usepackage[utf8]{inputenc}
\usepackage[colorlinks=true,citecolor=blue,linkcolor=blue]{hyperref}
\usepackage[normalem]{ulem}
\usepackage{amsmath,amssymb, mathrsfs}
\usepackage{epsfig}
\usepackage{graphicx}               
\usepackage{color}
\usepackage{slashed}
\usepackage{multirow}
\usepackage{placeins}
\usepackage[dvipsnames]{xcolor}
\usepackage{epstopdf}
\usepackage{soul}
\usepackage{tikz}
\usepackage[capitalise, english]{cleveref}
\usepackage{siunitx}
\usepackage{xspace}
\usepackage{booktabs}
\usetikzlibrary{trees}
\usetikzlibrary{decorations.pathmorphing}
\usetikzlibrary{decorations.markings}
\usepackage{aas_macros}
\usepackage{subfigure}

\newcommand\myshade{80}
\colorlet{mylinkcolor}{ForestGreen}
\colorlet{mycitecolor}{Aquamarine}
\colorlet{myurlcolor}{violet}

\hypersetup{
  linkcolor  = mylinkcolor!\myshade!black,
  citecolor  = mycitecolor!\myshade!black,
  urlcolor   = myurlcolor!\myshade!black,
  colorlinks = true
}

\definecolor{jblue}{RGB}{20,50,100}
\definecolor{npurple}{RGB} {153, 51, 204}
\definecolor{wred}{RGB}{217,0,56}
\definecolor{white}{RGB}{255,255,255}

\definecolor{korange}{RGB}{235, 80,  43}
\definecolor{korange2}{RGB}{245, 100,  63}
\definecolor{kyelloworange}{RGB}{255, 210,  110}
\definecolor{kyelloworange2}{RGB}{240, 170,  90}
\definecolor{kred}{RGB}{204,  102, 153}
\definecolor{kpurple}{RGB}{153,  61, 190}
\definecolor{kpurplelight}{RGB}{213,  161, 230}

 \definecolor{tobycolour}{rgb}{.5,.0,.5}

\DeclareSIUnit\year{yr}
\DeclareSIUnit\pc{pc}
\DeclareSIUnit\ergs{ergs}
\DeclareSIUnit\mSun{\ensuremath{M_\odot}}
\allowdisplaybreaks

\makeatletter
\providecommand*{\diff}%
  {\command{\lmultau}{\ensuremath{L_\mu-L_\tau}\xspace}
\new@ifnextchar^{\DIfF}{\DIfF^{}}}
\def\DIfF^#1{%
  \mathop{\mathrm{\mathstrut d}}%
    \nolimits^{#1}\gobblespace}
\def\gobblespace{%
  \futurelet\diffarg\opspace}
\def\opspace{%
  \let\DiffSpace\!%
  \ifx\diffarg(%
    \let\DiffSpace\relax
  \else
    \ifx\diffarg[%
      \let\DiffSpace\relax
    \else
        \ifx\diffarg\{%
        \let\DiffSpace\relax
      \fi\fi\fi\DiffSpace}

\usepackage{tikz,xcolor,hyperref}

\definecolor{lime}{HTML}{A6CE39}
\DeclareRobustCommand{\orcidicon}{\hspace{-1mm}
	\begin{tikzpicture}
	\draw[lime, fill=lime] (0,0) 
	circle [radius=0.16] 
	node[white] {{\fontfamily{qag}\selectfont \tiny \,ID}};
	\draw[white, fill=white] (-0.0525,0.095) 
	circle [radius=0.007];
	\end{tikzpicture}
	\hspace{-3mm}
}

\foreach \x in {A, ..., Z}{\expandafter\xdef\csname orcid\x\endcsname{\noexpand\href{https://orcid.org/\csname orcidauthor\x\endcsname}
			{\noexpand\orcidicon}}
}

\definecolor{aquamarine}{rgb}{0.2,0.7,0.6}
\definecolor{cerulean}{RGB}{0,166,214} 
\definecolor{subtlered}{rgb}{0.8,0.3,0.3}


\begin{document}

\title{$\nu$ limits from Super-Kamiokande on dark matter-electron scattering in the Sun}

\author{Dhashin Krishna\orcidA{}} \email{mdhashinkrishna@gmail.com} \affiliation{Centre for High Energy Physics, Indian Institute of Science, C.\,V.\,Raman Avenue, Bengaluru 560012, India}

\author{Rinchen Sherpa\orcidB{}}
\email{rinchens@iisc.ac.in}
\affiliation{Centre for High Energy Physics, Indian Institute of Science, C.\,V.\,Raman Avenue, Bengaluru 560012, India}

\author{Akash Kumar Saha\orcidC{}} \email{akashks@iisc.ac.in}
\affiliation{Centre for High Energy Physics, Indian Institute of Science, C.\,V.\,Raman Avenue, Bengaluru 560012, India}

\author{Tarak Nath Maity\orcidD{}}
\email{tarak.maity.physics@gmail.com}
\affiliation{School of Physics, The University of Sydney, ARC Centre of Excellence for Dark Matter Particle Physics, NSW 2006, Australia}

\author{Ranjan Laha\orcidE{}}
\email{ranjanlaha@iisc.ac.in}
\affiliation{Centre for High Energy Physics, Indian Institute of Science, C.\,V.\,Raman Avenue, Bengaluru 560012, India}

\author{Nirmal Raj\orcidF{}}
\email{nraj@iisc.ac.in}
\affiliation{Centre for High Energy Physics, Indian Institute of Science, C.\,V.\,Raman Avenue, Bengaluru 560012, India}

\date{\today}


\begin{abstract}

Particle dark matter scattering on electrons in the Sun may gravitationally capture and self-annihilate inside it to neutrinos and anti-neutrinos, or other final states that in turn decay to them.
Using up-to-date measurements by Super-Kamiokande of the fluxes of atmospheric electron-type and muon-type neutrinos, we set the most stringent limits on the electron scattering cross sections of dark matter down to about $10^{-40}-10^{-39}$~cm$^2$ over a mass range of 4$-$200\,GeV.
These outdo direct searches for dark matter-electron scattering and previously set limits at IceCube.
We also derive corresponding reaches at Hyper-K, and show that atmospheric neutrino observations restricted to the direction of the Sun can improve sensitivities. 
 
\end{abstract}

\maketitle

\section{Introduction}
\label{sec:Introduction}
     
While evidence for dark matter (DM) abounds in astrophysical and cosmological settings, detection of its non-gravitational interactions would mark an epoch in the hunt for its identity~\cite{Cirelli:2024ssz}.
One strategy is to look for the signals of particle DM scattering on Standard Model (SM) particles in terrestrial detector targets~\cite{Goodman:1984dc,DelNobile:2021wmp,Akerib:2022ort,Cooley:2021rws}.
In particular, searches for DM scattering on electrons have resulted in stringent null results~\cite{Kopp:2009et, Essig:2011nj, Essig:2015cda, XENON:2019gfn, SuperCDMS:2018mne, DarkSide:2018ppu, DarkSide-50:2022hin, DAMIC:2019dcn, EDELWEISS:2020fxc, Crisler:2018gci,SENSEI:2020dpa, PandaX:2022xqx, DAMIC-M:2023gxo, SENSEI:2023zdf, Dror:2020czw,  PandaX-II:2021nsg}.

Another attractive possibility for probing DM-SM scattering interactions, garnering much attention in recent literature, is the use of celestial objects as large-volume detectors.
This includes 
main-sequence stars counting the Sun~\cite{IceCube:2011aj,Bernal:2012qh,IceCube:2016dgk,IceCube:2021xzo,Gupta:2022lws,Bose:2021cou,Maity:2023rez, Chen:2023fgr, Bhattacharya:2024pmp,John:2024thz,Bose:2024wsh,Gupta:2025jte,Chu:2024gpe,Leane:2024bvh}, planets~\cite{Chauhan:2016joa,IceCube:2024yaw,Bramante:2022pmn,Renzi:2023pkn,Garani:2019rcb,Leane:2021tjj,French:2022ccb,Blanco:2023qgi,Robles:2024tdh,Bramante:2019fhi,Adler:2008ky,Blanco:2024lqw,Leane:2020wob,Bhattacharjee:2022lts,Linden:2024uph},
and Fermi-degenerate compact stars~\cite{Bramante:2023djs,Baryakhtar:2022hbu,Baryakhtar:2017dbj,Raj:2017wrv,Acevedo:2019agu,Joglekar:2019vzy, Joglekar:2020liw,Garani:2020wge, Busoni:2021zoe,Maity:2021fxw,Bramante:2021dyx, Bose:2021yhz,Bell:2023ysh,Bhattacharjee:2023qfi,Bhattacharjee:2024pis,Graham:2018efk,Acevedo:2019gre,Bell:2021fye,Acevedo:2023xnu,Coffey:2022eav,Garani:2023esk,Raj:2023azx,Raj:2024kjq,Basumatary:2024uwo,Bramante:2024ikc}.
One well-studied means by which these bodies probe DM interactions is through their gravitational capture of ambient DM following energy loss by scattering on their constituent particles.
The captured DM can either heat the body, form a destructive black hole, or self-annihilate to SM states that escape the celestial object to be detected at terrestrial experiments.
In this work, we study the capture of DM in the Sun via scattering on electrons, followed by annihilations to final states that result in a neutrino flux detectable at the Super-Kamiokande experiment as an excess over atmospheric neutrino events.
Using this we set the most stringent limits to date on this scenario.
Previous limits on DM-solar electron scattering using neutrinos at Super-K and IceCube may be found in Refs.~\cite{Kopp:2009et, Maity:2023rez}. 

\begin{figure*}
    \centering
    \includegraphics[width = \columnwidth]{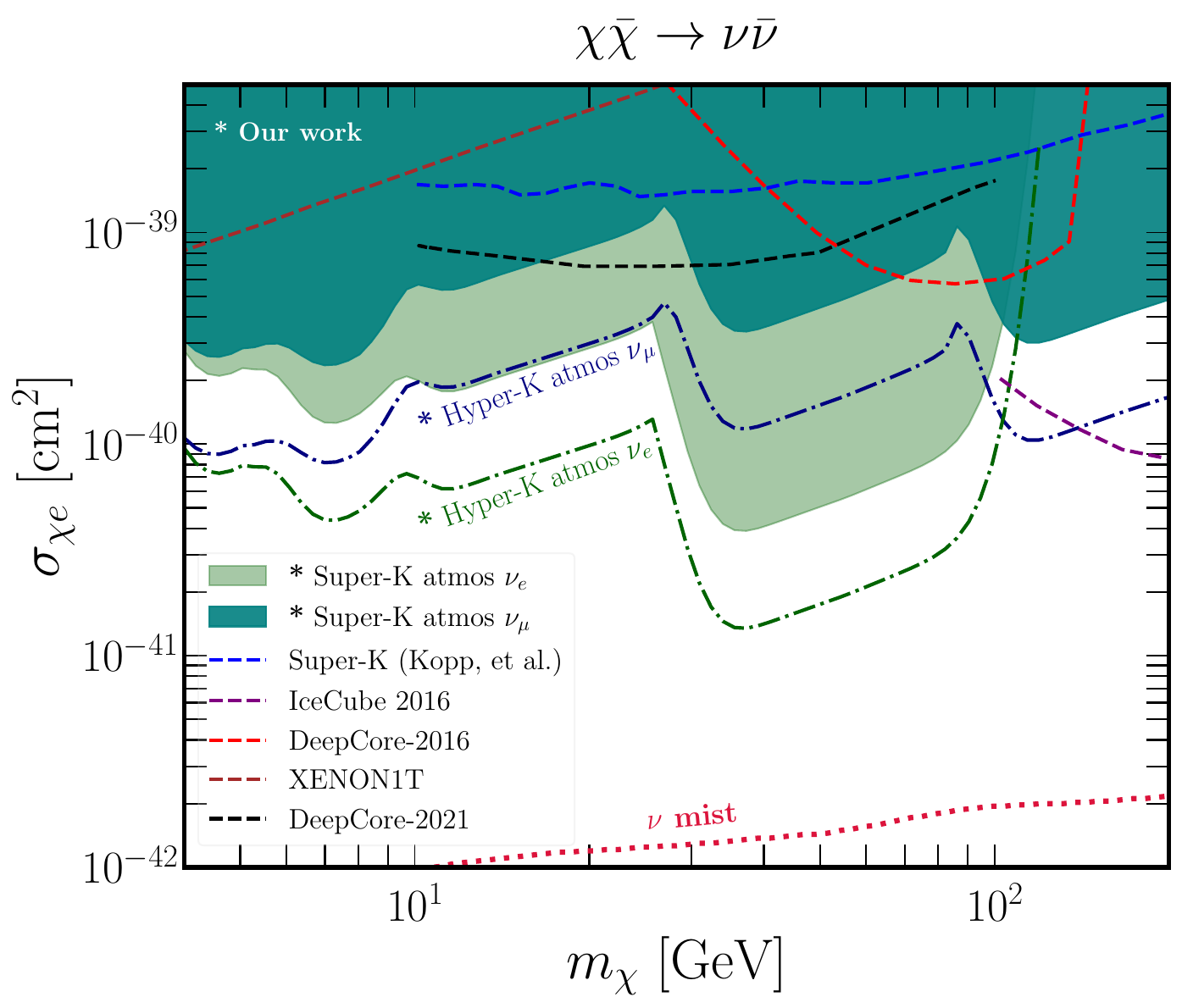}
    \includegraphics[width = \columnwidth]{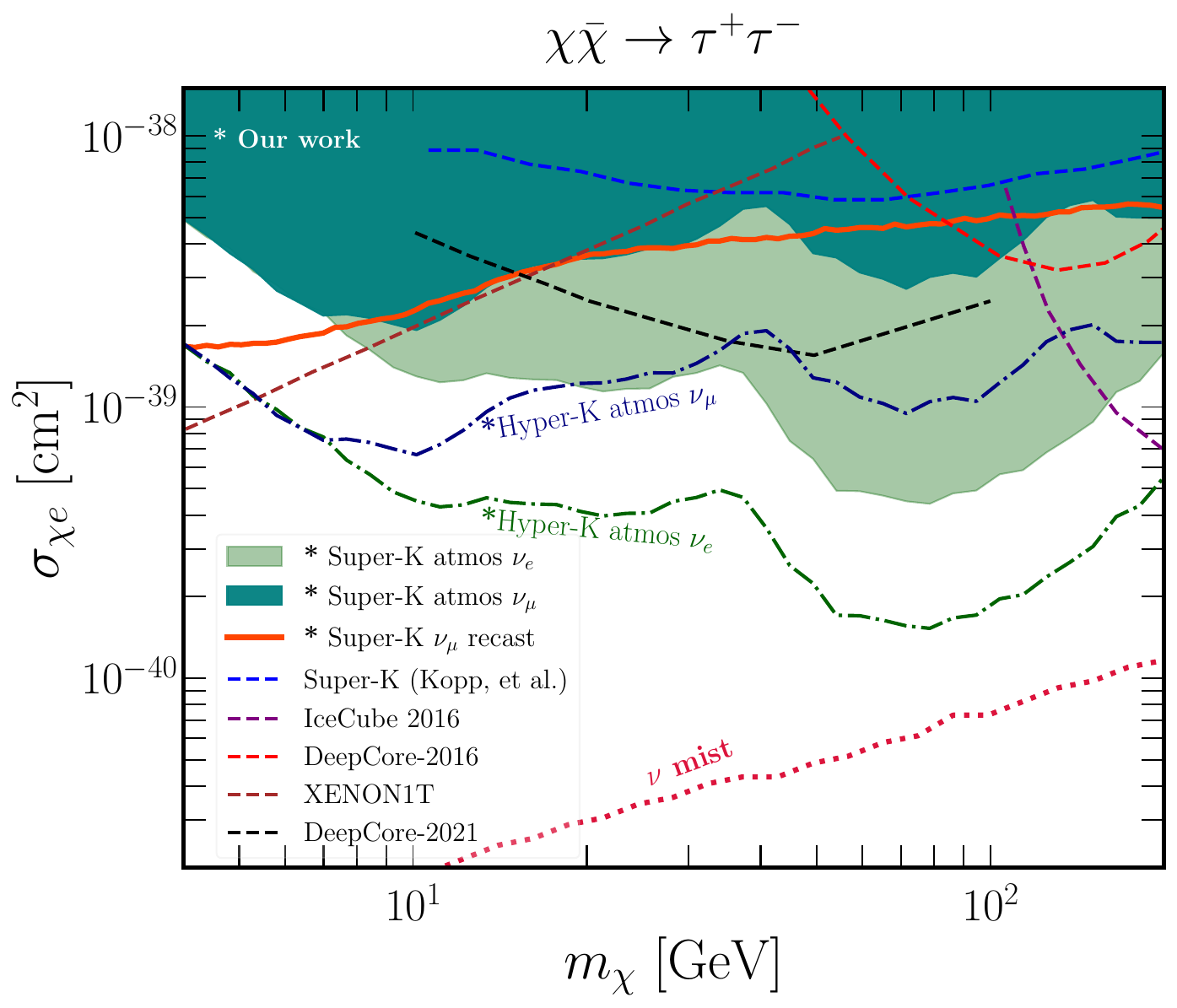}  
    \caption{Limits on DM-electron scattering cross section as a function of DM mass, assuming DM annihilations in the Sun to 
    $\nu \bar\nu$ on the {\bf left}, and to $\tau^+\tau^-$ (in turn decaying to $\nu_{e, \mu,\tau}$) on the {\bf right}.
     Shown here are our bounds from Super-K observations of atmospheric $\nu_e$ and $\nu_\mu$ fluxes\,\cite{Super-Kamiokande:2015qek} and corresponding future sensitivities at Hyper-K. 
    Also shown in the $\tau^+ \tau^-$ panel are our limits obtained from upper limits on the total integrated flux of $\nu_\mu$ from DM annihilations in the Sun in Ref.~\cite{Super-Kamiokande:2015xms}. 
    For comparison are shown limits from DeepCore-2021, DeepCore-2016, and IceCube-2016 observations of neutrinos from the solar direction~\cite{Maity:2023rez}, an older Super-K atmospheric neutrino dataset~\cite{Kopp:2009et}, and XENON1T direct searches for DM-electron scattering~\cite{XENON:2019gfn}. 
    The region below the dotted curve denotes the ``neutrino mist" where solar atmospheric neutrinos become a major background~\cite{Ng:2017aur}. 
    Note that the dynamical ranges of the y-axis are different across the panels.
    See Sec.~\ref{sec:results} for more details.}
    \label{fig:tautaununu}
\end{figure*}

Our main result is summarized in Fig.~\ref{fig:tautaununu}, which we will describe in detail in the rest of the work.
The key takeaway is that Super-K atmospheric $\nu$ datasets set the best bounds on the DM-electron scattering cross section over a wide range of DM masses.
In this way our study makes significant headway in probing models of so-called leptophilic DM~\cite{Krauss:2002px, Baltz:2002we, Ma:2006km, Hambye:2006zn, Bernabei:2007gr, Cirelli:2008pk, Chen:2008dh, Bi:2009md, Cao:2009yy, Goh:2009wg, Bi:2009uj, Ibarra:2009bm, Davoudiasl:2009dg,Dedes:2009bk, Kopp:2009et, Cohen:2009fz, Chun:2009zx, Chao:2010mp, Haba:2010ag,Ko:2010at, Carone:2011iw, Schmidt:2012yg, Das:2013jca, Dev:2013hka, Kopp:2014tsa, Chang:2014tea, Agrawal:2014ufa, Bell:2014tta, Freitas:2014jla,Cao:2014cda, Boucenna:2015tra,Lu:2016ups, Chauhan:2016joa, Garani:2017jcj, Duan:2017pkq, Duan:2017qwj,Chao:2017emq,Li:2017tmd, Ghorbani:2017cey,Sui:2017qra,Han:2017ars,  Madge:2018gfl,Junius:2019dci, Bell:2019pyc, YaserAyazi:2019psw, Joglekar:2019vzy,  Ghosh:2020fdc, Joglekar:2020liw, Chakraborti:2020zxt, Bell:2020lmm, Horigome:2021qof, Garani:2021ysl,Dasgupta:2020mqg, Bose:2023yll, Bhattacharya:2023stq,Asadi:2024jiy}. 
In the following sections we briefly review the mechanisms of DM solar capture, annihilations to $\nu$ fluxes, and Super-K measurements, before discussing our results and the scope of our study. 
Technical details are collected in the Supplemental Material.

\section{Set-up}
\label{sec:setup}

\subsection{Dark matter capture and annihilation in the Sun}
\label{subsec:dmcapann}

For a DM particle of any spin, mass $m_{\chi}$, and cross section for elastic scattering on electrons $\sigma_{\chi e}$, the rate of capture in the Sun is given by~\cite{Garani:2017jcj}
\begin{eqnarray}
\label{eq:cweak}
 C_\odot &=& 4\pi \int^{R_\odot}_0{dr  \ r^2 }\int^\infty_0{du_{\chi}\left(\frac{\rho_{\chi}}{m_{\chi}}\right)}\frac{f_{v_\odot}\left(u_\chi\right)}{u_\chi}w(r) \nonumber\\
&& \times \int^{v_{\rm esc}(r)}_0 dv R^-_e\left(w\rightarrow v\right),
\end{eqnarray}
where $R_\odot$ is the radius of the Sun, $\rho_\chi=0.3$ GeV/cm$^3$ is the solar neighborhood DM density, and $f_{v_\odot}(u_\chi)$ is the DM speed distribution in the solar frame. 
We denote the DM speed at infinity  as $u_\chi$, so that at a distance $r$ from the Sun the speed $w(r)=\sqrt{u_\chi^2+v_{\rm esc}^2(r)}$ for solar escape speed $v_{\rm esc}$. 
 We take from Ref.~\cite{Garani:2017jcj} the differential scattering rate $R^-_e$ that encodes the probability that a DM particle with speed $w$ scatters to reach a final speed $v<w$, with the AGSS09 model~\cite{Vinyoles:2016djt} used for solar electron number density and temperature profiles. 
Note that DM couplings to electrons would generically give rise to DM-nucleon scattering at the loop level, which may be a sub-dominant effect~\cite{Kopp:2009et,Bell:2019pyc}. 
We do not consider the model-dependent effect of nucleon scattering in this work and leave to future investigation the exploration of its possible interplay with electron scattering.

If the captured DM self-annihilates, the DM population in the solar interior may attain a steady state within the age of the Sun due to equilibrium between capture and annihilation rates.
This is insensitive to the annihilation cross sections above certain $m_\chi$-dependent values~\cite{Maity:2023rez}, and is easily achieved near thermal freeze-out values as allowed by indirect searches. 
We therefore assume this capture-annihilation equilibrium.
Then ignoring the effects of evaporation of the captured DM, as is valid for $m_\chi \gtrsim 4$ GeV~\cite{Garani:2021feo}, the annihilation rate of DM in the Sun is simply\,\cite{Peter:2009mk, Kopp:2009et}
\begin{eqnarray}
    \Gamma^{\rm ann}_\odot=\frac{C_\odot}{2}~.
    \label{eq:annirate}
\end{eqnarray}
In this work we consider the leptonic channels $\chi \bar\chi \to \nu\bar\nu$, where by $\nu \bar{\nu}$ we denote the average over all 3 flavors, and $\chi \bar\chi \to \tau^+ \tau^-$, which would further decay to produce neutrino final states.
The resultant energy flux of neutrinos on the Earth's surface is then
\begin{eqnarray}
    E_{\nu}^2 \frac{d \phi_{\nu}}{d E_{\nu}} = \frac{\Gamma^{\rm ann}_{\odot}}{4 \, \pi \, D_{\odot}^2} \times  E_{\nu}^2 \, \frac{dN_{\nu}}{dE_{\nu}},
\label{eq:neutrino_flux}
\end{eqnarray}
where $D_{\odot}$ is the Earth-Sun distance and $dN_\nu$/$dE_\nu$ is the spectrum of the final state neutrinos. 
Here we assume neutrinos of energies $\lesssim$ 500 GeV as otherwise their scattering cross sections would be so high as to have them significantly attenuated by the solar medium. 
As the neutrino energy is bounded from above by the DM mass, we consider only the conservative range of $m_\chi \leq 200$~GeV. 
For $\chi \bar \chi \to \nu \bar{\nu}$, the neutrino spectrum is a line feature at $m_\chi$, which will be broadened at Super-K due to its finite energy resolution.
 To account for this we Gaussian-smear the neutrino flux with a 10\% resolution as appropriate for our energies of interest~\cite{Shiozawa:1999sd,Drakopoulou:2017apf,Super-Kamiokande:2019gzr}. 
 In the $\chi \bar \chi \to \tau^+ \tau^-$ channel,
the $\tau^\pm$ decay daughters $\mu^{\pm}$, $K^{+}$ and $\pi^{+}$ may come to rest before decaying, with the mesons giving rise to monochromatic neutrino signals. 
We have used the package \texttt{$\chi$aro$\nu$}~\cite{Liu:2020ckq} that takes into account these interactions. 
We have checked that the $\tau^+ \tau^-$ channel produces neutrino spectra with both mono-energetic ``spikes" from stopped mesons (resulting in $\pi$ decay-at-rest and $K$ decay-at-rest peaks, seen for only $\nu_\mu$ spectra), and broad ``shoulders" from the prompt primary decay of mesons (seen for all $\nu$ flavors).
We compute the effects of neutrino propagation (including oscillations) in the Sun and vacuum using  \texttt{nuSQuIDS}~\cite{Arguelles:2021twb}, and in the atmosphere and rock of the Earth using \texttt{nuCraft}~\cite{Wallraff:2014qka}, with which we estimate the annual averaged neutrino oscillation probability.

\subsection{Super-K signals}
\label{subsec:superksignals}

\begin{figure*}
    \centering
  \includegraphics[width =\columnwidth]{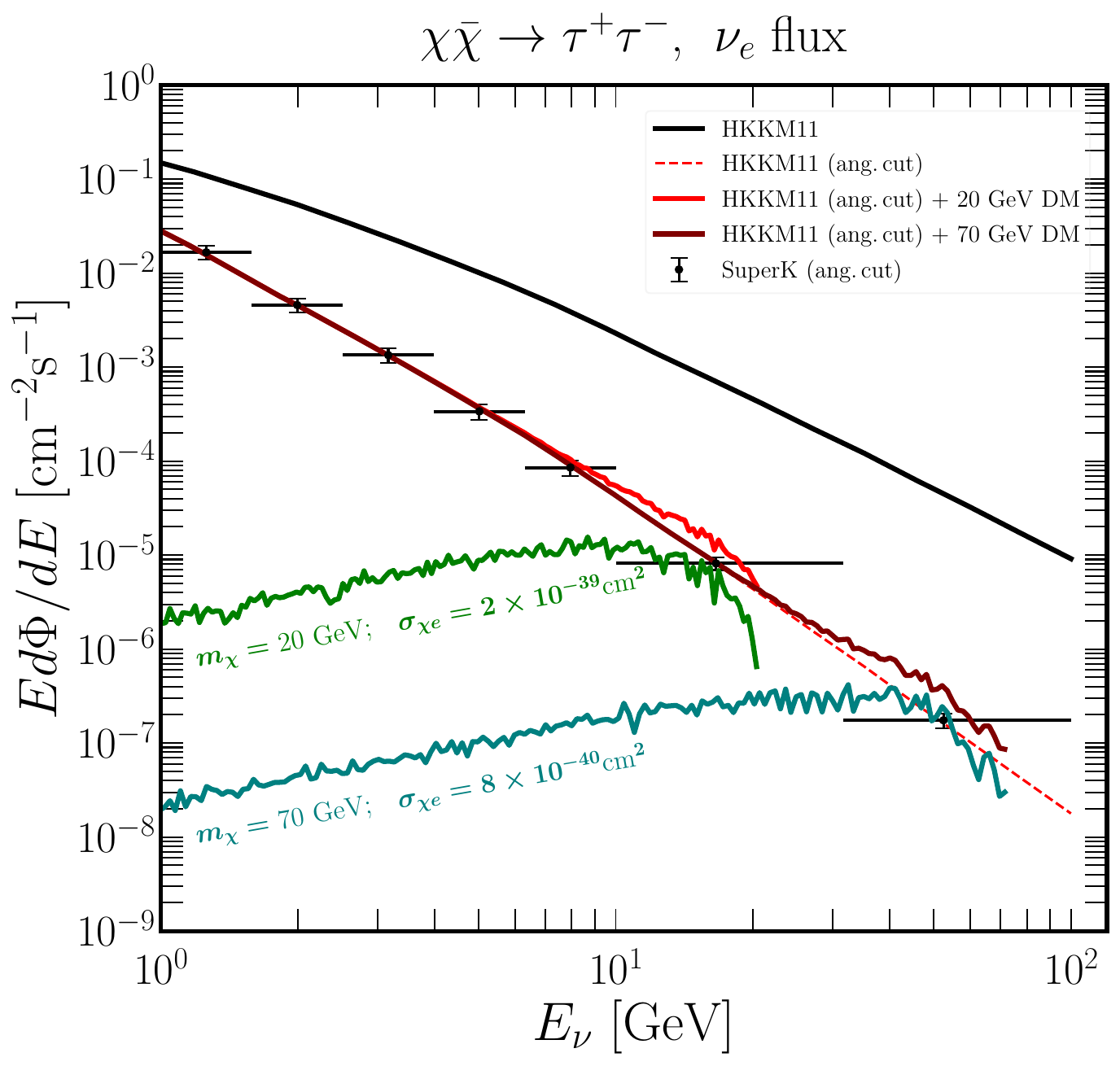} 
  \includegraphics[width =\columnwidth]{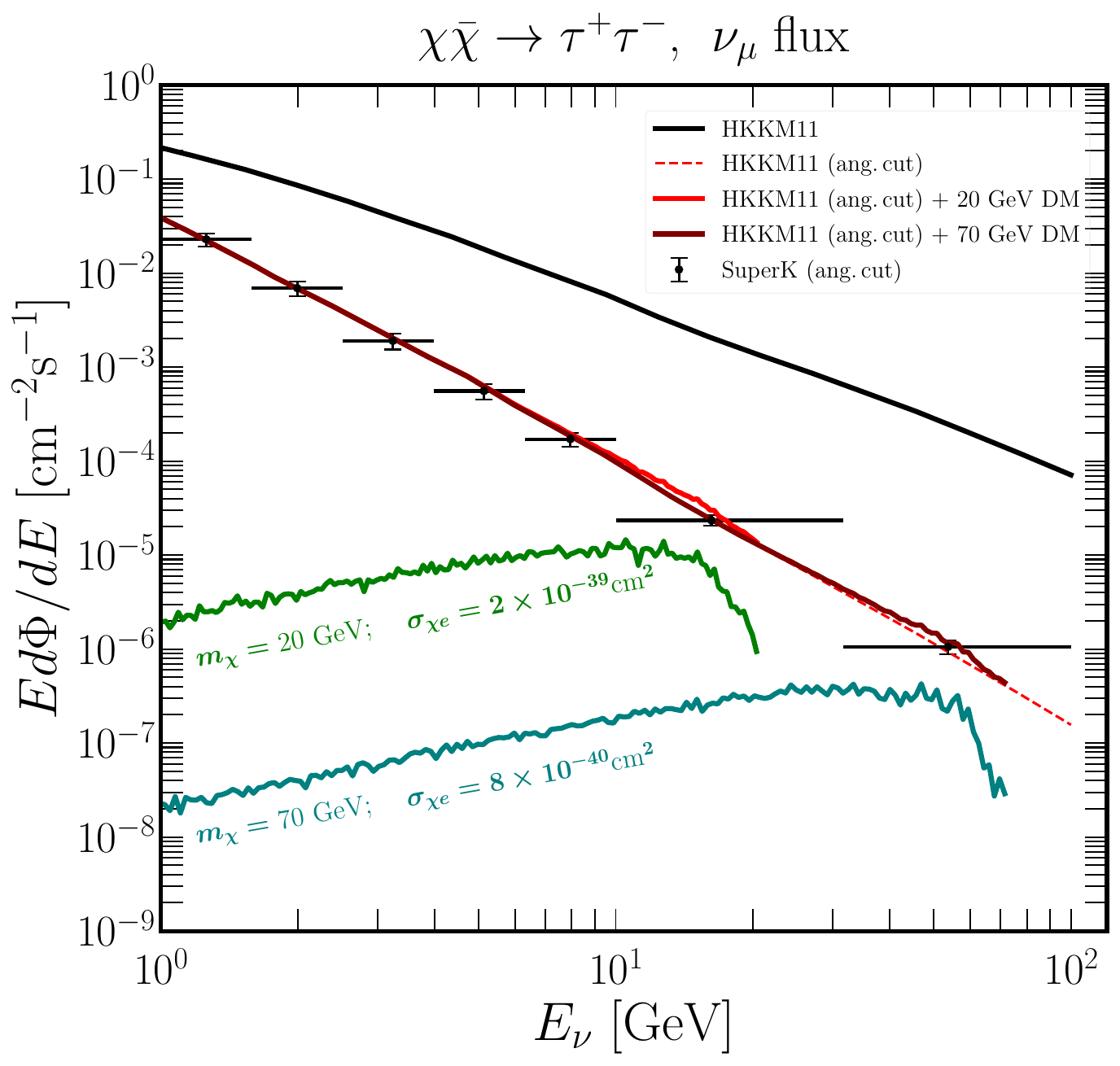}
      \caption{Atmospheric neutrino flux measured at Super-K, binned in energy and fit by the HKKM11 model, and the neutrino fluxes from $\chi \bar\chi \to  \tau^+ \tau^-$ in the Sun for the benchmark parameters indicated. 
 The DM signal combined with the atmospheric flux is seen to well exceed the measured flux near the energies where the signal peaks for $\nu_e$ data but only slightly for $\nu_\mu$ data.  
Thus these benchmarks are only ruled out by Super-K $\nu_e$ flux measurements as seen in Fig.~\ref{fig:tautaununu}.
See text for more details.}
        \label{fig:fluxnuenumu}
\end{figure*}

In Fig.\,\ref{fig:fluxnuenumu} we show an illustration of the flux of $\nu_e$ and $\nu_\mu$ obtained from DM annihilations in the Sun that may be detected by Super-K for $\{m_\chi, \sigma_{\chi e} \}$ of $\{20~{\rm GeV}, 2 \times 10^{-39}~{\rm cm}^2\}$ and  $\{70~{\rm GeV},  8 \times 10^{-40}~{\rm cm}^2\}$. 
For comparison are also shown Super-K data from Runs I$-$IV~\cite{Super-Kamiokande:2015qek} on the atmospheric $\nu$ flux and the HKKM11 model that fits it~\cite{Honda:2011nf}; this flux constitutes our main background.
In obtaining this plot, we have imposed an angular cut on the atmospheric $\nu$ flux, multiplying the per-steradian flux in Ref.\cite{Super-Kamiokande:2015qek} by 
\begin{equation}
2\pi \int_0^{\delta\theta_{\rm tot}(E_\nu)} \,d\theta\ {\rm sin}\theta~ = 4\pi \sin^2 \left(\frac{\delta\theta_{\rm tot}(E_\nu)}{2}\right)~,
\label{eq:angcut}
\end{equation}
where $\delta\theta_{\rm tot}(E_\nu)$  is the total angular resolution as a function of incident neutrino energy; see the Supplemental Material.  
This cut ensures that we only analyze the flux originating from the solar direction and discard the rest of the diffuse atmospheric $\nu$ background. 
For our benchmark parameters the combined fluxes of atmospheric and DM-sourced neutrinos are visibly in excess of Super-K $\nu_e$ measurements in Fig.~\ref{fig:fluxnuenumu}, and may therefore be constrained.
We note that these benchmarks are safe from XENON1T constraints as seen in Fig.~\ref{fig:tautaununu}.

\section{Results}
\label{sec:results}

In Fig.\,\ref{fig:tautaununu}, we show our Super-K limits on the DM-electron scattering cross section vs DM mass.
We also show the future reach of Hyper-K~\cite{Hyper-Kamiokande:2018ofw} via appropriately scaling the detector fiducial mass with respect to Super-K, so that the $\sigma_{\chi e}$ sensitivities are $\sqrt{187/22.5} \simeq 2.9$ times stronger.
To set these limits we first compute
 \begin{equation}
 \chi^2 = \sum_{i=1}^{N_{\rm bin}} \frac{\left(\phi_{\rm d}^i - \phi_{\rm atm}^i - \phi_{\chi}^i\right)^2}{{\sigma_{\rm d}^i}^2}~,  
 \label{eq:chisq}
 \end{equation}
where in the $i^{\rm th}$ bin $\phi_{\rm d}^{i}$ is the measured neutrino flux, $\sigma_{\rm d}^{i}$ its corresponding uncertainty, and $\phi_{\rm atm}^i$ \& $\phi_{\chi}^i$ are respectively the HKKM11 model \& DM signal average flux, and then
obtain the 95\% C.L. upper limits on $(\sigma_{\chi e},\, m_\chi)$ by setting  $\chi^2-\chi_{\rm min}^2 = 2.71$, where $\chi_{\rm min}^2$ is the minimum value of $\chi^2$ obtained by varying the DM parameters~\cite{Lamperstorfer:2015cfg}.
We use the same energy binning considered in Ref.~\cite{Super-Kamiokande:2015qek}.   
We consider energy bins up to $\pm$20\% of $m_\chi$ for the $\nu \bar \nu$ channel accounting for the detector energy resolution, and up to 1.2 $m_\chi$ for the  $\tau^+ \tau^-$ channel; in practice the resolution has negligible impact on the broad $E_\nu$ spectrum. 
In any case, due to the $\mathcal{O}(100\%)$ width of the energy bins (see Fig.~\ref{fig:fluxnuenumu}) the Super-K energy resolution 
has a negligible effect on the limits we derive.
For the $\tau^+ \tau^-$ channel we also present in our parameter space limits from a dedicated search at Super-K for a $\nu_\mu$ flux from DM annihilations in the Sun~\cite{Super-Kamiokande:2015xms}.
Finally, we also show the ``neutrino mist" region where our signal events $\leq$ events from an irreducible flux of solar atmospheric neutrinos, degrading sensitivities.
For this we have used the flux of the ``naive" model derived in Ref.~\cite{Ng:2017aur}.
The region shown denotes the $\nu_\mu$ mist, although we expect the $\nu_e$ mist sensitivities to be comparable.
Here we remark that values of $\sigma_{\chi e}$ at the edge of the neutrino mist region do not satisfy the capture-equilibrium condition that resulted in Eq.~\eqref{eq:annirate}, so we used the full solution for the annihilation rate as derived in, e.g., Ref.~\cite{Garani:2017jcj}.

It may be seen that we constrain new ranges of parameters beyond previous limits.
This is mainly because the large density and size of the Sun generically help in capturing DM with small scattering cross sections.
For both $\nu \bar{\nu}$ and $\tau^+\tau^-$ channels the $\nu_e$ data is seen to provide stronger limits on $\sigma_{\chi e}$ than the $\nu_\mu$ data, which is simply due to the smaller $\nu_e$ flux produced in the atmosphere. 
Here we note that the net angular resolution in detection, dominated by the scattering angle, does not make a difference in the limits between $\nu_e$ and $\nu_\mu$ as it is comparable for both channels; see the Supplemental Material.

Our limits in both panels of Fig.~\ref{fig:tautaununu} exhibit wiggle-like features due to the nature of the binning of Super-K data.
In particular, the signal neutrino flux peaks near the DM mass (Fig.~\ref{fig:fluxnuenumu}), and since only the flux up to the corresponding $E_\nu$ drives the limit in Eq.~\eqref{eq:chisq}, increasing $m_\chi$ while the edge of the signal spectrum stays within a corresponding $E_\nu$ bin weakens the constraints. 
However, when $m_\chi$ is further increased so that the signal spectrum reaches the next energy bin, the limits strengthen.

\section{Discussion}
\label{sec:discussion}

In this work we have set stringent, world-leading limits on DM-electron scattering using the most up-to-date measurements of atmospheric neutrinos at Super-K, assuming that DM captured in the Sun annihilates to $\nu \bar{\nu}$ or $\tau^+\tau^-$. 
Hyper-K will be able to extend these sensitivities by a factor of a few.
Despite their smaller volume than Hyper-K, other imminent neutrino telescopes such as DUNE\,\cite{Rott:2019stu,Chauhan:2023zuf} and JUNO\,\cite{Guo:2015hsy,JUNO:2023vyz} would also be interesting targets to study thanks to their complementary detection channels and technologies.
Other DM annihilation final states giving rise to neutrino fluxes may also be studied in these contexts, and DM masses higher than considered here may also be constrained by carefully treating the effects of attenuation of neutrinos in the Sun that become more important at high neutrino energies.
While we had applied a simple angular cut in Eq.~\eqref{eq:angcut} to select neutrinos from the solar direction, more robust limits may be obtained by using the angular dependence of atmospheric neutrinos or, as done in Ref.\,\cite{Super-Kamiokande:2015xms}, by applying the angular cut for all Super-K event categories.

Overall, we anticipate improved reaches with larger datasets and better understanding of detector systematics.\\

{\bf \em Note added.}

While this work was under progress, Ref.~\cite{Nguyen:2025ygc} appeared on arXiv investigating a scenario similar to ours.
The limits obtained in that work  on $\sigma_{\chi e}$ using $\nu_\mu$ detection at Super-K are seen to be stronger by up to an order of magnitude than ours near $m_\chi = 10$~GeV, and are comparable near $m_\chi = 100$~GeV.
We believe one source of this discrepancy is the DM solar capture rate obtained in Ref.~\cite{Nguyen:2025ygc}.
While Ref.~\cite{Nguyen:2025ygc} uses expressions for the capture rate derived in Ref.~\cite{Garani:2017jcj}, the capture rates plotted in Fig.~S3 exceed those plotted in Fig.~1 of Ref.~\cite{Garani:2017jcj} by a factor of about 7 for the relevant DM mass range.
Our capture rates match those in Ref.~\cite{Garani:2017jcj}.
Further, we also differ in our estimates of the angular cut used to reduce the atmospheric neutrino background.
Whereas we use Eq.~\eqref{eq:angcut}, Ref.~\cite{Nguyen:2025ygc} mentions following Ref.~\cite{Robles:2024tdh}, which has used the square of the net angular resolution that underestimates the backgrounds by roughly a factor of 3. 
The cut in Ref.~\cite{Nguyen:2025ygc} is stronger than ours also due to a difference in treatment of charged lepton scattering angles at detector events. 
Reference~\cite{Nguyen:2025ygc} takes the mean scattering angle as a function of neutrino energy from Refs.~\cite{Konishi:2010mv,Konishi:2011sc}, which only includes quasi-elastic neutrino-nucleus scattering. 
We determine the mean scattering angles by also including resonant and deep inelastic scattering as relevant for high neutrino energies that are of interest here.
Moreover, for setting limits Ref.~\cite{Nguyen:2025ygc} integrates over the flux obtained from the atmospheric model HKKM11 to estimate the total number of background events, and uses Poisson statistics to derive a constraint on the number of DM-induced signal events.
On the other hand, ours is a binned analysis that includes the data points and uncertainties in the Super-K measurements, which are dominated by systematics (see Fig.~6 of Ref.~\cite{Super-Kamiokande:2015qek}).
Finally, we note that we set limits using both the $\nu_e$ and $\nu_\mu$ fluxes measured at Super-K, with $\nu_e$ giving better limits, whereas Ref.~\cite{Nguyen:2025ygc} uses only the $\nu_\mu$ flux.

\section*{Acknowledgments} 
We thank Bhavesh Chauhan,  Marco Cirelli, Raghuveer Garani, Srubabati Goswami, Joachim Kopp, and Thomas Schwetz for useful discussions.
We also thank Marta Czurylo, Biprajit Mondal, Danilo Piparo, and Jonas Rembser for help with \texttt{ROOT}. 
R.S.\,\,acknowledges the University Grants Commission (UGC), Government of India, for financial support via the UGC-NET Senior Research Fellowship.  A.K.S.\,\,acknowledges the Ministry of Human Resource Development, Government of India, for financial support via the
Prime Ministers’ Research Fellowship (PMRF).
The work of T.N.M.\,\,is supported by the
Australian Research Council through the ARC Centre of Excellence for Dark Matter Particle Physics. R.L.\,\,acknowledges
financial support from the institute start-up funds and ISRO-IISc STC for the grant
no.\,\,ISTC/PHY/RL/499.

The first three authors contributed equally to this work.
\bibliography{refs.bib}

\clearpage
\newpage
\maketitle
\begin{center}
\textbf{\large Supplemental Material} 
\end{center}

\begin{figure}[h]
    \centering
    \includegraphics[width = \columnwidth]{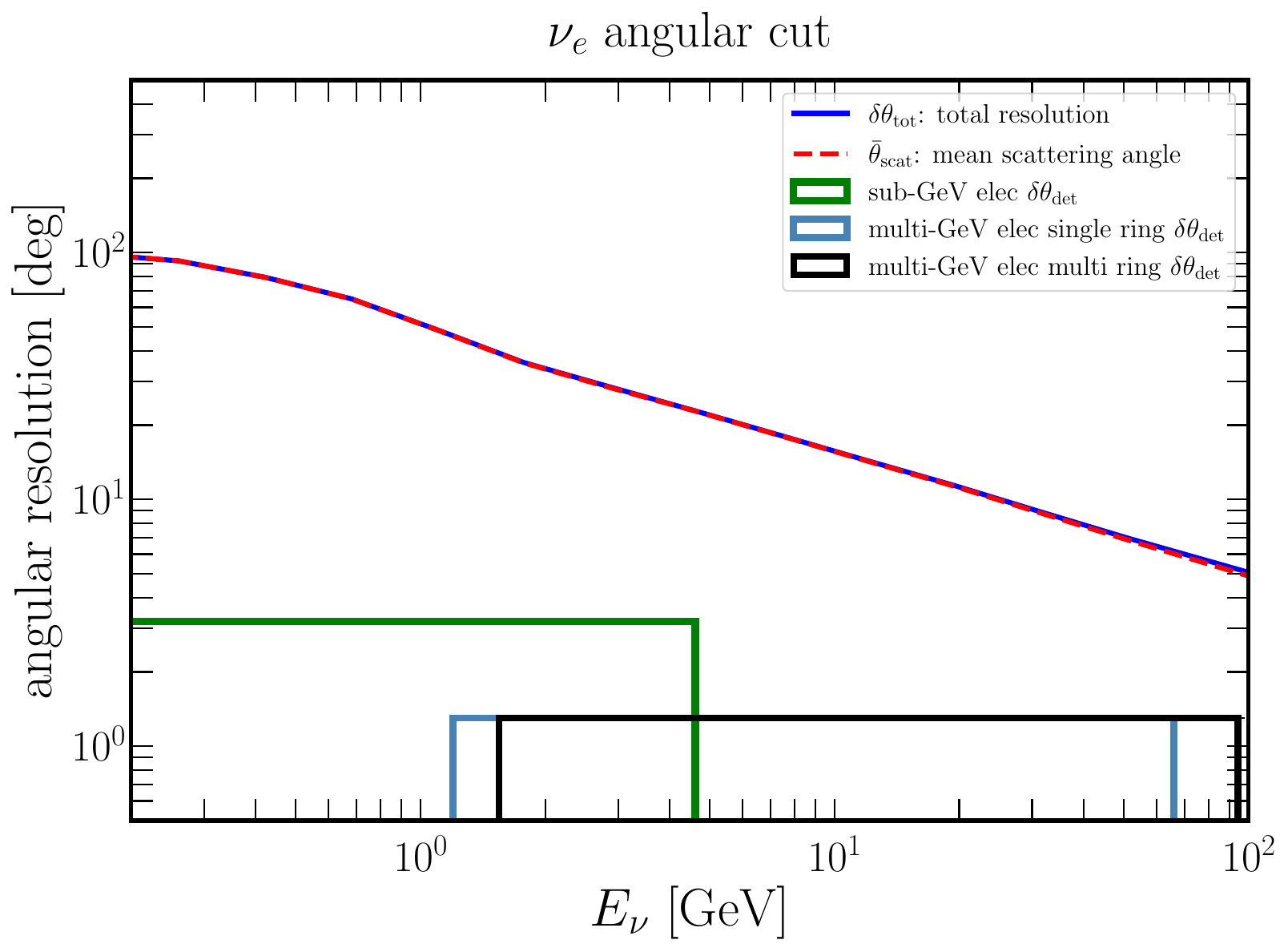}
\\ \includegraphics[width = \columnwidth]{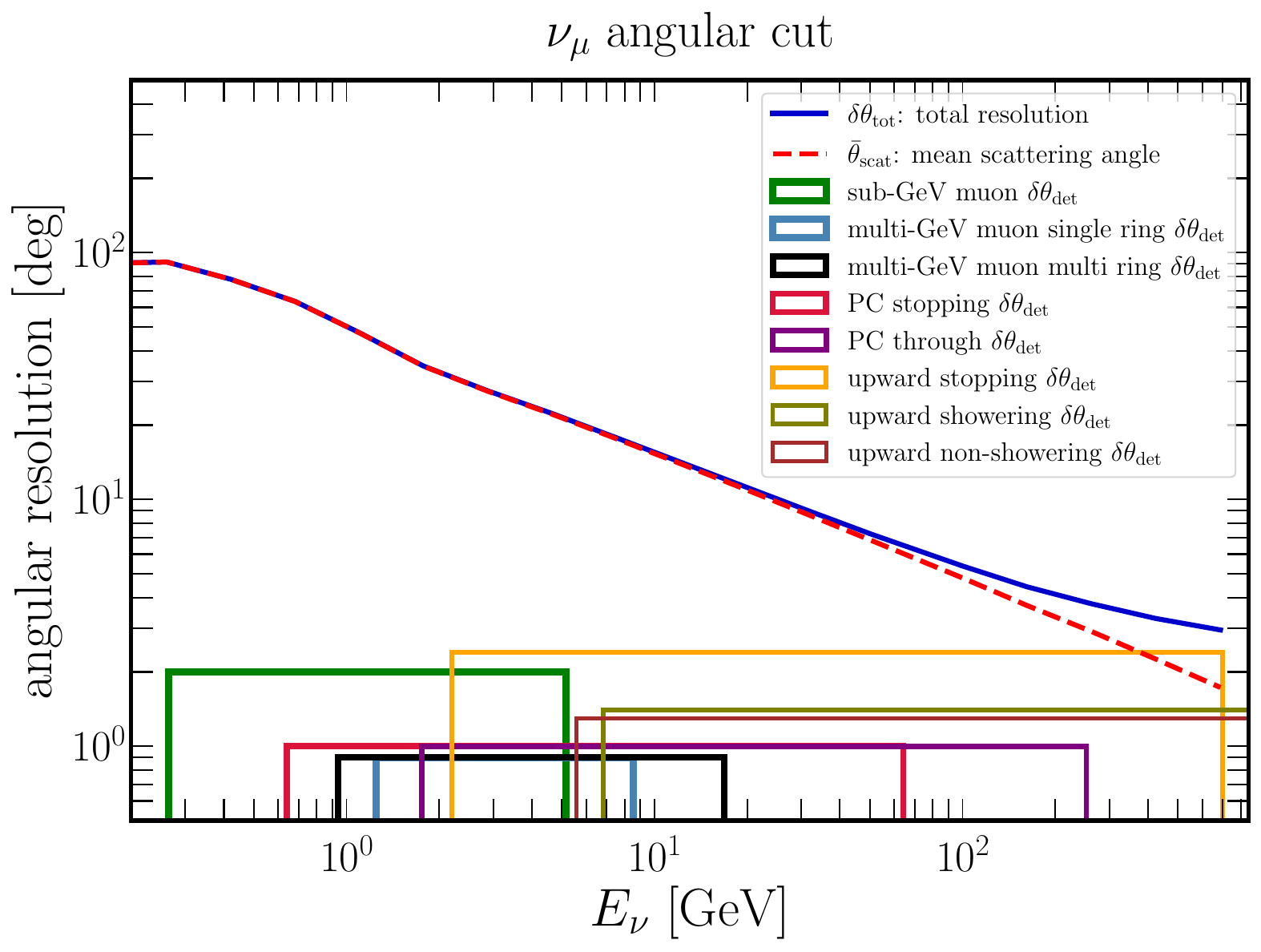}
    \caption{Angular uncertainties from the mean scattering angle of the outgoing charged lepton (obtained with \texttt{NuWro}) and detector resolutions of various event types at Super-K as described in Section~B of Supplemental Material. 
    The net angular resolution is seen to be dominated by the scattering angle except at high $\nu_\mu$ energies.}
    \label{fig:angres}
\end{figure}

\subsection{Charged current cross sections from NuWro}

We use the Monte Carlo neutrino event generator \texttt{NuWro}~\cite{Golan:2012wx} to obtain neutrino-H$_2$O charged current scattering cross sections and the average scattering angle of the final state $e^\pm/\mu^\pm$. 
In the parameter initialization file, we set 
\texttt{SF\_method}=1, and neutral current interactions to zero, and generate $10^5$ events for each neutrino energy. 
We checked that the total cross sections thus obtained match with the output of GENIE~\cite{Andreopoulos:2009rq}, and note that the cross sections for $\bar \nu$ are generally smaller than for $\nu$. \\

\subsection{Resolutions for the angular cut}

To impose the angular cut in Eq.~\eqref{eq:angcut} for reducing the atmospheric neutrino background, we must estimate the energy-dependent total angular resolution for neutrino-H$_2$O scattering events in Super-K, given by
\begin{equation}
    \delta \theta_{\rm tot} (E_\nu) = \sqrt{\bar\theta_{\rm scat}^2 (E_\nu) + \delta\theta_{\rm det}^2 (E_\nu)}~,
\end{equation}
where $\bar \theta_{\rm scat}$ is the mean scattering angle of the outgoing charged lepton and $\delta \theta_{\rm det}$ is the detector angular resolution. 
 We obtain $\bar \theta_{\rm scat}$ by generating $10^5$ events as described above. 
 As the mean scattering angle for $\nu_\mu$ ($\nu_e$) is larger than that of $\bar\nu_\mu$ ($\bar\nu_e$), we conservatively use the former.
 We also find that the the mean scattering angle dominates $\delta \theta_{\rm tot}$ except at the highest $\nu_\mu$ energies considered.
 Nonetheless, for the sake of completeness we describe here the determination of $\delta \theta_{\rm det}$.

The detector angular resolution depends on the type of event in Super-K, broadly classified as:

(i) {\bf Fully contained (FC) events} where the lepton is produced and contained in the inner detector (ID). 
FC events are further divided into 7 sub-GeV and 6 multi-GeV sub-types on the basis of the species of particle and number of Cherenkov rings produced. 
(ii) {\bf Partially contained (PC) events} in which the lepton is produced in the ID but deposits energy in the outer detector (OD). PC event particles can either pass through the OD (``OD through-going") or stop in the OD (``OD stopping").
(iii) {\bf Upward-going-muons (Up-$\mu$)} are selected to identify
neutrinos that interact with the rock surrounding the detector and produce muons. 
This is to mitigate the background of atmospheric muons.
Up-$\mu$ can again either stop in the detector (``upward stopping") or pass through it (``upward-through going"). 
Further, through-going Up-$\mu$ events can be either showering or non-showering~\cite{Choi:2015nxy}. 
We take the detector angular resolution for different event types from Refs.~\cite{Hagiwara:2020tkq} and 
\cite{Super-Kamiokande:2007uxr}. 
To obtain conservative estimates we choose the maximum $\delta \theta_{\rm det}$ possible at a given energy for any event type.
We depict the detector angular resolutions of these event types in Fig.~\ref{fig:angres}.

\vspace{0.05in}

\textbf{} \\

\vspace{0.07in}

\end{document}